\documentstyle[twocolumn,aps,fleqn,epsfig,epsf,graphicx,amssymb,latexsym]{revtex}
\newcommand{\R}{{\cal{R}}}
\def\la{\mathrel{\mathchoice {\vcenter{\offinterlineskip\halign{\hfil
$\displaystyle##$\hfil\cr<\cr\sim\cr}}}
{\vcenter{\offinterlineskip\halign{\hfil$\textstyle##$\hfil\cr
<\cr\sim\cr}}}
{\vcenter{\offinterlineskip\halign{\hfil$\scriptstyle##$\hfil\cr
<\cr\sim\cr}}}
{\vcenter{\offinterlineskip\halign{\hfil$\scriptscriptstyle##$\hfil\cr
<\cr\sim\cr}}}}}

\def\vec#1{\ifmmode\mathchoice{\mbox{\boldmath$\displaystyle#1$}}
{\mbox{\boldmath$\textstyle#1$}}
{\mbox{\boldmath$\scriptstyle#1$}}
{\mbox{\boldmath$\scriptscriptstyle#1$}}\else
\hbox{\boldmath$\textstyle#1$}\fi}
\begin{document}
\title{Absence of supersensitivity to small input signals in generalized on--off
systems}
\author{Eurico Covas\thanks{Web: http://www.eurico.web.com} and Reza Tavakol\thanks{Email: reza@maths.qmw.ac.uk}}
\address{Astronomy Unit, Mathematical Sciences, Queen Mary \& Westfield College, Mile End
Road, London, United Kingdom}
\date{\today}
\twocolumn
\maketitle
\begin{abstract}
It has recently been shown
that nonlinear skew product
dynamical systems with invariant subspaces which are capable of displaying
on--off intermittency can show supersensitivity to small input signals.

Here we show that this supersensitivity is absent for more general
dynamical systems with non--skew product structure, 
capable of displaying a generalized 
form of on--off intermittency, and is
therefore in this sense fragile. 
This absence of supersensitivity
is of importance in view of the fact that
dynamical systems are generically
expected to be of non--skew product
nature.
\end{abstract}
Many dynamical systems of interest possess symmetries or constraints
which force the presence of invariant subspaces. A great deal of
effort has recently gone into the study of such systems
(see e.g.\ \cite{pikowsky1984,plattetal1993,ashwinetal1996}).
A sub--class of these models, namely those with skew product
structure (and normal parameters\footnote{Parameters that leave the
dynamics on the invariant manifold unchanged are called normal,
otherwise they are referred to as non--normal.}),
have been shown to be capable of producing a number of novel
modes of behavior, including on--off intermittency
\cite{plattetal1993} and bubbling
\cite{ashwinetal1996}.

Recently, Zhou and Lai \cite{zhouetal1999} have
shown that systems of this type can display supersensitivity, in the
sense that small constant or time varying inputs to the system can
induce extremely large responses. The authors further claim that with an
additional odd symmetry condition, this supersensitivity is robust to
addition of noise. Such supersensitivity could be
of importance in many fields, including the study of 
synchronization of coupled chaotic systems \cite{pecoraetal1990}
and the design of sensor devices \cite{bohmeetal1994}.

The results on on--off intermittent systems reported by these
authors can all be described within the framework
of skew product systems.
Generically, however, one would expect typical dynamical
systems to have non--skew product structure (with non--normal parameters).
Systems of this type have recently been studied and have
been found to be capable of displaying a number of additional novel dynamical modes of
behavior,
absent in skew product systems,
including a generalization of on--off intermittency, referred to as
{\em in--out
intermittency} \cite{ashwinetal1999}.

The easiest way to characterise in--out
intermittency is by contrasting it with on--off intermittency. Briefly, let
$M_I$ be the invariant subspace and $A$ the attractor which exhibits either
on--off or in--out intermittency. If the intersection $A_0=A\cap M_I$ is a
minimal attractor, then we have on--off intermittency, whereas if $A_0$ is not
a minimal attractor, then we have in--out intermittency. In the latter case
there can be different invariant sets in $A_0$ associated with attraction and
repulsion transverse to $A_0$, hence the name in--out. Another crucial
difference between the two is that, as opposed to on--off intermittency, in the
case of in--out intermittency the minimal attractors in the invariant
subspaces do not necessarily need to be chaotic and hence the trajectories
can (and often do) shadow a periodic orbit in the `out' phases
\cite{ashwinetal1999}.
\\

Our aim here is to find out whether
this type of supersensitivity, observed in
on--off intermittent systems,
persists in more general
non--skew product systems
which are capable of displaying in--out intermittent behaviour.

A simple class of maps that can model
both on--off and in--out types of intermittency is given by
\begin{equation} \label{general-map}
x_{n+1}=F(x_n,y_n, {\bf a}),\ y_{n+1}=G(x_n,y_n, {\bf a}),
\end{equation}
where $G(x_n, 0, {\bf a})=0$, the variables $x_n$ and $y_n$ represent the
dynamics within the invariant submanifold ($y=0$) and the transverse distance to it
respectively and ${\bf a \in \R^m}$ are the control parameters of the
system.
A special subset of these systems, with skew product form over
the dynamics in $x$, can be written as
\begin{equation}\label{eqskewprod}
x_{n+1}=F(x_n,{\bf a}),\ y_{n+1}=G(x_n,y_n, {\bf a}).
\end{equation}
By considering a skew product system of type (\ref{eqskewprod}),
Zhou and Lai \cite{zhouetal1999} modelled the motion
near the invariant submanifold $y=0$, using
a Fokker--Planck equation.
In this way they were able to predict that the
sensitivity $S$ of the map in the neighbourhood of
a blowout bifurcation, leading to on--off intermittency, is given by
\begin{equation}\label{sensitivity-on-off}
S=\frac{\langle y\rangle}{p}=\frac{\tau}{p\ln(\tau/p)},
\end{equation}
where $\langle y\rangle$ is the average of the transverse variable $y$, $p$ is the
input signal and $\tau$ is the threshold below which $y$ goes through a laminar phase.
They were able to
confirm this prediction numerically.

To study whether non--skew product (in--out intermittent) systems
can also display supersensitivity,
we considered a particular example of the map (\ref{general-map})
in the form
\begin{eqnarray}\label{eqmap}
x_{n+1}&=& r x_n(1-x_n)+s x_n y_n^2,\\
y_{n+1}&=& \nu e^{bx_n}y_n+ay_n^3\nonumber,
\end{eqnarray}
where $r\in[0,4]$ and $(s,\nu,a,b)\in\R$ are the control parameters.
Note that for $s=0$, the map (\ref{eqmap}) has the skew product form
(\ref{eqskewprod}) and for fixed $r$, the parameters $s$, $a$, $b$ or $\nu$
are normal.
Thus depending upon the choice of its parameters,
this map is capable of displaying
both on--off and in--out types of intermittency,
Note also 
that this map
possesses the odd symmetry condition
$G(-y) = - G(y)$ that was
found in \cite{zhouetal1999} to
be required for the
robustness of supersensitivity with respect
to noise. Also the transverse
Lyapunov exponent $\lambda_T$ for this map can be readily calculated
to be
\begin{equation}\label{lyt}
\lambda_T=\ln\nu +b\langle x\rangle_r,
\end{equation}
where $\langle x\rangle_r$ is the average of the variable $x_n$ for an initial condition
on the invariant submanifold $y=0$. 

To study the effects of
an input signal on in--out systems, we considered
a variant of this map given by
\begin{eqnarray}\label{inputmap}
x_{n+1}&=& r x_n(1-x_n)+s x_n y_n^2,\\
y_{n+1}&=& \nu e^{bx_n}y_n+ay_n^3 +p\nonumber,
\end{eqnarray}
where the real parameter $p$ models the effects of a small input
signal.

\begin{figure}[!htb]
\centerline{\def\epsfsize#1#2{0.42#1}\epsffile{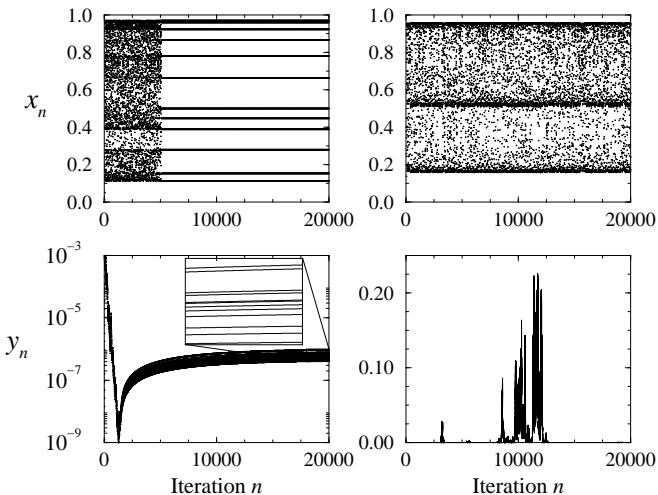}}
\caption{\label{series.p=1e-10}
Comparative study of bursting behaviour
in in--out (left panels) and on--off (right panels) regimes, for
an input signal $p=10^{-10}$.
Note how in--out dynamics (lower left panel) is insensitive to the
input signal, in comparison with on--off (lower right panel).
The  upper left panel and the inset on
the lower left panel also demonstrate clearly the
presence of the period 12 attractor in
the invariant submanifold.
The parameters values are
$r=3.8800045$, $\nu=1.82$, $b=-1.020625$, $a=-1$ and $s=-0.3$ for the in--out
case and $r=3.82786$, $\nu=1.82$, $b=-1.006$, $a=-1$ and $s=0$ for the on--off
case.}
\end{figure}

To begin with, we made a comparative numerical study of
the sensitivity of in--out and on--off systems to input signals
$p$, using (\ref{inputmap}). Fig.\ \ref{series.p=1e-10} shows a
comparative study of the bursting behaviour in
the two cases close to, but below, their blowout points.
As can be seen from the comparison of
the lower panels, there is very little bursting in the in--out case.

\begin{figure}[!htb]
\centerline{\def\epsfsize#1#2{0.48#1}\epsffile{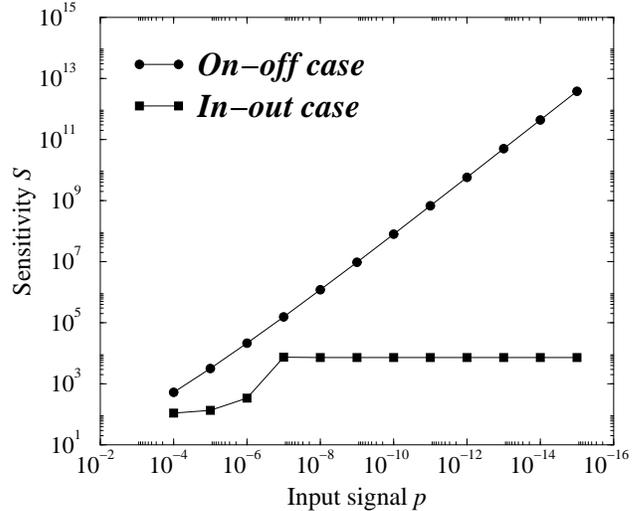}}
\caption{\label{sensitivity.versus.p}
Dependence of the sensitivity $S$ on the input signal $p$
for in--out and on--off cases.
Note the relative insensitivity of the in--out
to the input signal and the saturation in sensitivity
in this case.
The parameter values are as
in Fig.\ \ref{series.p=1e-10}.}
\end{figure}

To further demonstrate this relative insensitivity in the in--out case,
we made a study of the sensitivity $S$ of the systems close to their blowout
points,
as a function of the input signal $p$. This is shown in Fig.\ \ref{sensitivity.versus.p},
which again demonstrates a distinct absence of supersensitivity for the in--out case,
specially for the lower input signal levels. Furthermore, it shows
a saturation in sensitivity in the in--out case for input signals
$p\la 10^{-7}$.

These results indicate a clear lack of supersensitivity
in the in--out case to small constant input signals.

To understand
this qualitative difference between the on--off and
in--out cases, we briefly recall a number of
differences between the two cases, relevant to our
discussion here. In the case of on--off,
the attraction and the ejection of the orbits near the
invariant manifold are brought about by a single chaotic attractor in $M_I$. Thus
for the values of the control parameter close to but below the blowout
point, the chaotic attractor in $M_I$ becomes transversally weakly
attracting, but there can be repelling orbits within this attractor that
are transversally unstable, leading to bubbling and allowing the orbit
to access the lower and upper boundaries frequently. The system thus
becomes sensitive to small inputs, producing large bursts and
hence supersensitivity.

\begin{figure}[!htb]
\centerline{\def\epsfsize#1#2{1#1}\epsffile{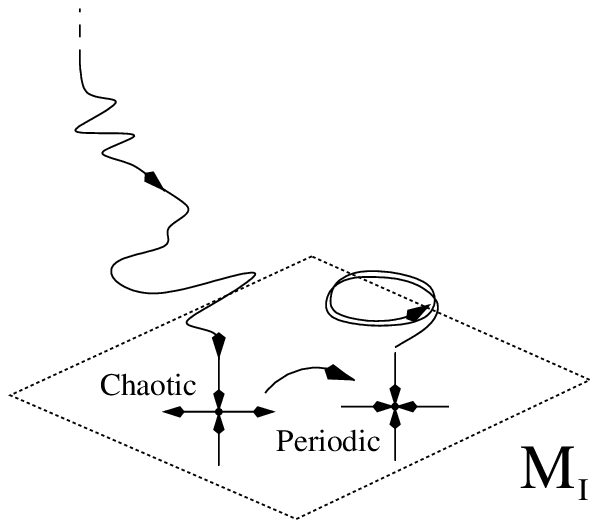}}
\caption{\label{after.blowout}
Schematic diagram showing the main dynamical features of the in--out
process near the blowout point with $\lambda_T < 0$.
This picture
does not qualitatively change in presence of a
input signal $p$, even though the period 12 attractor and the 
chaotic saddle are
slightly shifted from the previous $M_I$ (represented by the dotted line).}
\end{figure}

For the in--out case, on the other hand, the `in' and `out' phases are
governed by two separate invariant sets in $M_I$: a chaotic saddle and
a periodic attractor respectively. Thus for the values of control parameter ($b$ in
our case) above the blowout value, the chaotic saddle in the invariant
submanifold is transversally attracting whereas the periodic
attractor in $M_I$ is transversally unstable with a positive transverse
Lyapunov exponent $\lambda_T$. As a result, an orbit drawn towards the
invariant submanifold by the chaotic saddle is thus ejected by the
transversally unstable periodic attractor, leading to in--out
intermittency.
On the other hand, for the values of control parameter
$b$ just below the blowout value, the
unique periodic attractor in the
invariant submanifold becomes transversally stable
(with $\lambda_T <0$), while the chaotic saddle
still remains transversally
attracting. As a result orbits drawn towards
the invariant submanifold by the chaotic saddle
get attracted to the
periodic orbit there (see Fig.\ \ref{after.blowout} for a schematic
depiction).

We shall now show that
the presence of a small input
signal $p$ leaves the above dynamical picture
essentially
unaltered, apart from displacing the location of the
periodic attractor off the previous invariant
submanifold.
There are two ways to see this.
Firstly, for small values of the input
signal $p\ll {\cal O}(1)$,
the periodic orbit is expected to persist by continuity.
We numerically confirmed that
the period 12 orbit
involved in the in--out intermittency studied here (see \cite{ashwinetal1999}
for details)
does indeed survive for small values of $p$,
albeit shifted slightly off the invariant
submanifold $M_I$ (see the left panels of Fig.\ \ref{series.p=1e-10}).

Alternatively,
we can estimate the transverse location of the
displaced periodic orbit. To do this we recall that
we are interested in small displacements from $M_I$,
which implies that as a first approximation
we may ignore higher order dependence on $y$.
We therefore approximate the map (\ref{inputmap})
by
\begin{eqnarray} \label{linear-map}
x_{n+1}&=&rx_n (1- x_n)+s x_ny_n^2,\\
y_{n+1}&\approx& \nu e^{b x_{n}} y_{n} + p, \nonumber
\end{eqnarray}
where the second order term in $y$ has been kept
in the $x$ map in order to ensure the essential overall
non--skew product structure of the system.

The period 12 attractor involved in this case
has $x_{n}$ values
satisfying $x_{n+12} = x_{n}$ and $y_n$ values given by
\begin{eqnarray*}
y_0&=&p,\nonumber\\
y_1&=&\left(\nu e^{b x_0}+1\right) p,\nonumber\\
y_2&=&\left(\nu^2 e^{b (x_0+x_1)}+\nu e^{b x_1}+1\right) p,\nonumber\\
&\vdots&\nonumber\\
y_{12}&=&\left(\nu^{12} e^{b (x_0+x_1+\ldots+x_{11})}+\ldots+\nu e^{b x_{11}}+1\right)
p,\nonumber\\
y_{13}&=&\left(\nu^{13} e^{b (2 x_0+x_1+\ldots+x_{11})}\right.\nonumber\\
&&\left.+\nu^{12} e^{b (x_0+x_1+\ldots+x_{11})}+\ldots+\nu e^{b x_0}+1\right) p,\nonumber\\
&\vdots&\nonumber\\
y_n&=&\sum_{j=0}^{n} \left( \nu^j e^{b\left[
\sum\limits_{i=1}^{12}
\left\lfloor\frac{j+11-(i mod 12)}{12}\right\rfloor
x_{(i+n-1) mod 12}
\right]} \right) p,
\end{eqnarray*}
where $\lfloor \ \rfloor$ denotes the integer part.

The above expressions for $y_n$ change periodically (with period 12),
depending on the initial $x$. The asymptotic average
value of $y$ can then be approximated by
\begin{eqnarray}\label{yn_average}
\lim_{n \to \infty } \langle y\rangle &\approx&\sum_{j=0}^{n} \left( \nu^j e^{b\left[
\sum\limits_{i=0}^{11}
\frac{j}{12}
x_{i}
\right]} \right) p,\nonumber\\
&=&\left(
1-\nu e^{\frac{b}{12} \sum\limits_{i=0}^{11} x_{i}}
\right)^{-1}p\nonumber\\
&=&\frac{p}{1-\nu e^{\lambda_{T}-\ln \nu}}=
\frac{p}{1-e^{\lambda_T}},
\end{eqnarray}
where we have used (\ref{lyt}).
Interestingly this enables us to
find the sensitivity $S$ as a function of $\lambda_T$
\begin{equation}\label{sensitivity_analytical}
S=\frac{1}{1-e^{\lambda_T}},
\end{equation}
which is independent of $p$, thus explaining the saturation
in sensitivity $S$ observed in the in--out case in Fig.\
\ref{sensitivity.versus.p},
in clear contrast to expression for the 
on--off sensitivity given by (\ref{sensitivity-on-off}) .

It now remains to show that apart from the above shift
off the submanifold, the
periodic attractor remains essentially
intact.
To see this,
recall that
the effect of a non--zero $p$ on
$x$ is given by 
\begin{equation}
x_{n+1}\sim F^n(x_1, {\bf a})+s x_n {\langle y\rangle}^2,
\end{equation}
where $F^n(x_1, {\bf a})$ represents the $x$ component of the n$^{\mathrm th}$ iterate
of the map (\ref{eqskewprod}). Using (\ref{yn_average})
this  gives
\begin{equation}
|s| x_n {\langle y\rangle}^2 \sim |s|\, {\cal O}(1)\left[\frac{p}{1-e^{\lambda_T}}\right]^2.
\end{equation}
Now for input signals $p\ll {\cal O}(1)$ and
for the parameter regimes
chosen here,
$1-e^{\lambda_T}\gg \sqrt{p}$,
which implies
\begin{equation}
|s| x_n {\langle y\rangle}^2 \ll p,
\end{equation}
showing that to this approximation the $p$-induced variations in $x$ are
extremely small (relative to $p$), hence providing a good indication that
the periodic attractor remains essentially
intact.

\begin{figure}[!htb]
\centerline{\def\epsfsize#1#2{0.46#1}\epsffile{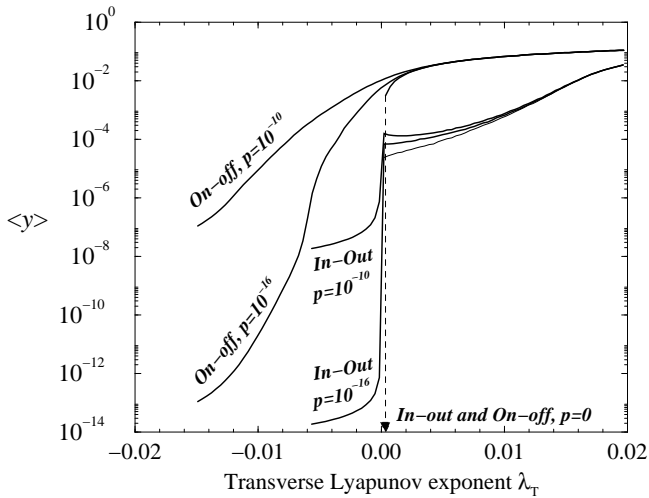}}
\caption{\label{average_y.versus.lyt}
Dependence of the average blowout variable $\langle y \rangle$
as a function of $\lambda_T$,
for
fixed input signals.
The parameters values are
as in Fig.\ \ref{series.p=1e-10}.}
\end{figure}

The above arguments and results demonstrate
the qualitative differences between
the responses of the on--off and the in--out dynamics 
to small input signals.
In particular, the survival of the
periodic orbit in the
latter case acts to
trap the incoming orbits and
therefore blocks the possibility
of supersensitivity in
this case. We expect this picture to be
common and thus supersensitivity
to be absent in the generic non--skew product (in--out) settings.

\begin{figure}[!htb]
\centerline{\def\epsfsize#1#2{0.46#1}\epsffile{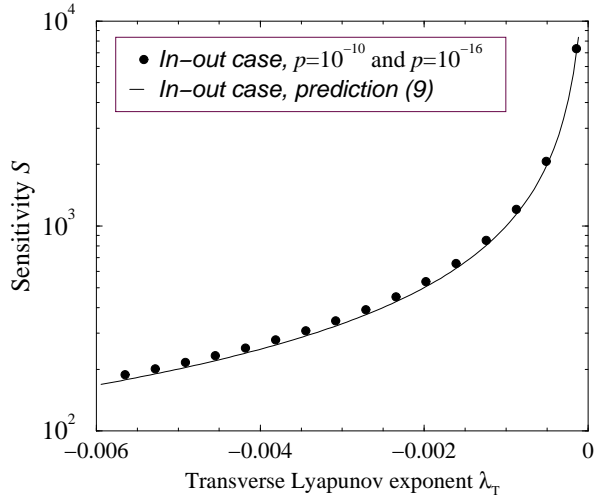}}
\caption{\label{sensitivity.versus.lyt.comparison}
Dependence of the sensitivity $S$ for a small input signal $p$
as a function of $\lambda_T$, together with 
the predicted scaling (\ref{sensitivity_analytical}). The parameter values are as
in Fig.\ \ref{series.p=1e-10}.}
\end{figure}

To further substantiate this finding,
we calculated the average
blowout variable $\langle y \rangle$ in system (\ref{inputmap}) for
fixed input signals,
as a function of $\lambda_T$.
The results are summarised in Fig.\ \ref{average_y.versus.lyt},
which show that for $\lambda_T>0$, both cases are relatively
insensitive to input signals, whereas for $\lambda_T<0$, the on--off
case is much more sensitive to input signals than
the in--out case, with the latter dependence 
in very good agreement with our prediction (\ref{yn_average}).

Finally we calculated the dependence of the
sensitivity $S$ for the in--out case as a function of $\lambda_T$,
with different input signals,
as a function of $\lambda_T$.
The results are shown in Fig.\ \ref{sensitivity.versus.lyt.comparison},
together with our predicted
expression (\ref{sensitivity_analytical}),
which show excellent agreement.
\\

To summarise, we have argued that the supersensitivity found
in \cite{zhouetal1999} for the case of skew product systems with
on--off intermittency is absent in the more general setting
of non--skew product systems, capable of displaying in--out
intermittency. We have substantiated this claim both analytically
and through
extensive numerical simulations.
We have also checked that the absence of supersensitivity in the in--out
case remains
robust to changes in both the input signal (of the form $p\sin(2\pi x)$)
as well as to unbiased noise in the transverse direction
(of the order of the input signal).

The absence of supersensitivity for systems displaying
in--out intermittency is important, particularly given
that dynamical systems are generically expected to be of non--skew product type.

\vspace{0.1cm}
EC is supported by a PPARC fellowship and RT
benefited from PPARC UK Grant No.\ L39094.

\end{document}